\documentclass{ieeeaccessarxiv}
\usepackage{caption}
\usepackage{cite}
\usepackage{amsmath,amssymb,amsfonts}
\usepackage{algorithmic}
\usepackage{graphicx}
\usepackage{textcomp}
\usepackage[hidelinks]{hyperref}
\def\BibTeX{{\rm B\kern-.05em{\sc i\kern-.025em b}\kern-.08em
    T\kern-.1667em\lower.7ex\hbox{E}\kern-.125emX}}

\newif\iftr
\newif\ifconf
\trtrue
\conffalse

\usepackage{url}

\begin{document}

\newcommand{\maciej}[1]{\textcolor{blue}{[Maciej: #1]}}
\newcommand{\raghu}[1]{\textcolor{blue}{[Raghu: #1]}}
\newcommand{\robert}[1]{\textcolor{blue}{[Robert: #1]}}
\newcommand{\greg}[1]{\textcolor{blue}{[Greg: #1]}}

\title{Hardware Acceleration for Knowledge Graph Processing: Challenges \& Recent Developments}
\iftr
\author{
\uppercase{Maciej Besta}\authorrefmark{1},
\uppercase{Robert Gerstenberger}\authorrefmark{1},
\uppercase{Patrick Iff}\authorrefmark{1},
\uppercase{Pournima Sonawane}\authorrefmark{2},
\uppercase{Juan Gómez Luna}\authorrefmark{1},
\uppercase{Raghavendra Kanakagiri}\authorrefmark{3},
\uppercase{Rui Min}\authorrefmark{4},
\uppercase{Grzegorz Kwa\'{s}niewski}\authorrefmark{1},
\uppercase{Onur Mutlu}\authorrefmark{1},
\uppercase{Torsten Hoefler}\authorrefmark{1},
\uppercase{Raja Appuswamy}\authorrefmark{5},
\uppercase{Aidan O Mahony}\authorrefmark{2}
\address[1]{ETH Zurich, Zürich, Switzerland}
\address[2]{Dell Technologies, Ovens, Co. Cork, Ireland}
\address[3]{Indian Institute of Technology Tirupati}
\address[4]{HIRO-MicroDataCenters BV, Heerlen, Netherlands}
\address[5]{Eurecom, Biot, France}
}
\fi

\iftr
\markboth
{Besta \headeretal: Hardware Acceleration for Knowledge Graph Processing: Challenges \& Recent Developments}
{Besta \headeretal: Hardware Acceleration for Knowledge Graph Processing: Challenges \& Recent Developments}
\fi

\iftr
\corresp{Corresponding authors: maciej.besta@inf.ethz.ch, Aidan.Omahony@dell.com}
\fi

\begin{abstract}
Knowledge graphs (KGs) have achieved significant attention in recent years, particularly in the area of the Semantic Web as well as gaining popularity in other application domains such as data mining and search engines. Simultaneously, there has been enormous progress in the development of different types of heterogeneous hardware, impacting the way KGs are processed. The aim of this paper is to provide a systematic literature review of knowledge graph hardware acceleration. For this, we present a classification of the primary areas in knowledge graph technology that harnesses different hardware units for accelerating certain knowledge graph functionalities. We then extensively describe respective works, focusing on how KG related schemes harness modern hardware accelerators. Based on our review, we identify various research gaps and future exploratory directions that are anticipated to be of significant value both for academics and industry practitioners.
\end{abstract}

\begin{keywords}
Knowledge Graphs, Semantic Web, Hardware Architectures, Systematic Literature Review, Graph Algorithms, Heterogeneous Hardware, FPGA, GPU, ASIC, CPU
\end{keywords}

\titlepgskip=-15pt

\maketitle

\section{Introduction}
\label{sec:introduction}
\PARstart{K}{nowledge} graphs (KG) are structured representations of information that are used to represent and organize data in a way that is easily accessible and understandable. They are used in a variety of applications, including information retrieval, natural language processing, and artificial intelligence (AI)~\cite{reinanda2020knowledge, venkatesh2022conversational, schneider2022decade, melluso2022enhancing, tiddi2020knowledge, schramm2024comprehensible}.

In the current era of data-centric ecosystems, it has become vitally important
to organize and represent the enormous volume of knowledge appropriately.
Recently, knowledge graphs have risen as a powerful tool for representing complex associations among entities and concepts across various domains, enhancing semantic search. As these knowledge graphs continue to grow in both scale and complexity, conventional computing methods encounter difficulties in effectively processing and analysing them in real-time, and addressing these challenges has prompted the investigation of hardware acceleration as a potential solution.

Hardware acceleration involves harnessing the capabilities of
specialized hardware components designed to perform specific tasks more
efficiently than what would be possible using a general-purpose Central
Processing Unit (CPU), such as Graphics Processing Units (GPUs) or Field
Programmable Gate Arrays (FPGAs). Both GPUs and FPGAs~\cite{blogpost2023a,
blogpost2023b} use strategies such as optimized memory use and low-precision
arithmetic to accelerate computation, adding a boost to CPU server engines. While hardware acceleration has
demonstrated remarkable success in various computational domains, its impact on
knowledge graph processing remains relatively unexplored. Hardware
acceleration has the potential to substantially enhance the performance of knowledge graph applications, enabling quicker and more precise data processing and analysis.

In this systematic literature review, we will examine the applications and consequences of hardware acceleration on knowledge graphs. We will review the existing literature on this topic and identify the main findings and trends in the use of hardware acceleration in knowledge graph applications. We will also consider the potential benefits and drawbacks of hardware acceleration, and we identify challenges and opportunities for future research and development.
%
%

\section{Overview of Knowledge Graphs}

Knowledge Graphs (KGs) accumulate and convey knowledge of the real world. They
can effectively organize data to represent complex information, so that it can
be efficiently and extensively explored in traditional and advanced
applications, offering significant benefits for data exploitation in creating
new knowledge. Knowledge graphs have emerged as an approach for the systematic
representation of knowledge of real-world entities in a machine-readable format~\cite{kg_acm_survey}.

\subsection{Representations}

A knowledge graph is usually modeled using one of two data
representations~\cite{besta2023demystifying}: the Labeled Property Graph (LPG),
also called property graph, and the Resource Description Framework
(RDF). The \textbf{LPG model} categorizes vertices and edges with the help of labels and
allows attributes for vertices and edges in the form of key-value pairs as
properties. \textbf{RDF} represents knowledge graphs in the form of triples, where each
triple consists of a subject, a predicate, and an object. Formally, edges are represented as triples $(h, r, t)$, where $h$ and $t$ are the head
and tail entities, and $r$ is the relation between them.

Knowledge graphs, regardless of the used data model, capture the relationships
between entities and are composed of entities (vertices, also referred to as nodes) and relationships
(edges), forming a graph structure that allows complex interconnections and
associations to be easily visualized and understood.


\subsection{Embeddings}

The knowledge graph input is typically human-readable, however certain tasks
benefit from a transformed (embedded) machine representation.
\textbf{Knowledge Graph Embedding (KGE)} models provide a way to represent entities (vertices) and relations (edges) of a knowledge graph in vector spaces, referred to as embeddings. These models capture the semantics of the graph and are used in various downstream tasks like link prediction, classification, and recommendation.

The goal of training a
knowledge graph embedding model is to learn embeddings for entities and relations such that the embeddings of the head and tail entities are close to each other in the embedding space when connected by a relation. This is achieved
by optimizing a loss function that captures the likelihood of the observed triples in the graph.
One common approach is to train on both positive and negative triples. A positive triple is an observed triple in the graph, and a negative triple is a corrupted version of a positive triple. The negative triples are sampled by replacing the head or tail entity of a positive triple with another entity in the graph.

\subsection{Benefits and Applications}

The benefits of knowledge graphs are manifold. They provide a structured and semantically rich representation of knowledge, which can be leveraged for various applications. For instance, KGs can be used to improve search engine results by understanding the context and semantics behind a user's query. They can also be used in recommendation systems to provide more personalized and context-aware recommendations~\cite{kg_recommender_survey}.

In addition to these, knowledge graphs have been successfully applied in
numerous other domains. For instance, in the pharmaceutical industry, KGs can be
used to represent complex relationships between drugs, diseases, and patients,
thereby aiding in drug discovery and personalized medicine~\cite{dgl-ke,
moliere}. In the field of social sciences, knowledge graphs can be used to analyze social networks and understand the dynamics of social interactions~\cite{sheth2019knowledge, qian2016anonymizing, he2020constructing}.

In education, several knowledge graph-based applications focus on supporting
remote teaching and learning. For example, considering the importance of course
allocation tasks in universities, a knowledge graph-based approach was proposed to
automate this task. One could construct a course knowledge graph in which the entities are courses, lecturers, course books, and authors in order to suggest relevant courses to students~\cite{nair2022knowledge, liang2023construction, albreiki2024clustering, grevisse2018knowledge}.


In healthcare, the growth of the medical sector has led to more options for
treatments. To help with this, medical recommender systems, especially
biomedical knowledge graph-based recommender systems (such as doctor and
medicine recommender systems), have been developed.  For instance, in
recommending medications, one can construct a heterogeneous graph whose nodes
are medicines, diseases, and patients to recommend accurate and safe medicine
prescriptions for patients with complicated medical issues~\cite{teng2020explainable, zhang2022traditional, rotmensch2017learning, gatta2017generating, cope2022maps, li2020real, wang2022application, qu2022review, chandak2023building}.

\begin{figure*}[t]
  \centering
  \includegraphics[width=1.0\linewidth]{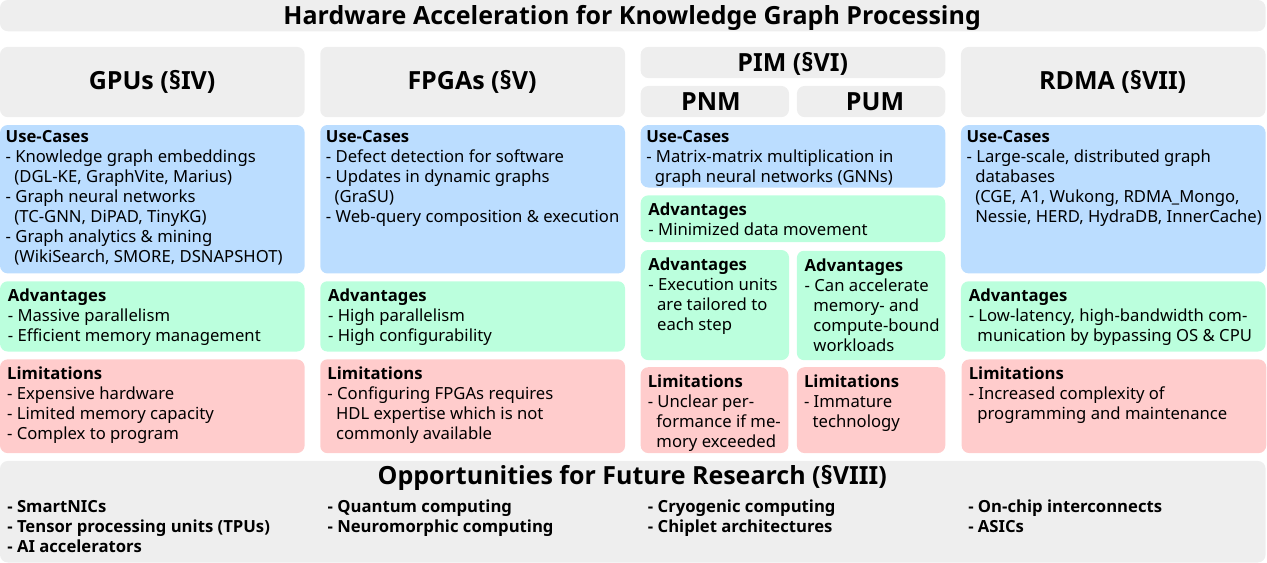}
  \caption{Overview of different hardware acceleration areas used for knowledge graph processing.}
  \label{fig:overview}
\end{figure*}

Various works also propose to enhance
general generative models with knowledge graphs.
The focus of these works
is usually to use KGs in order to enhance the LLM answers, for example by
grounding knowledge in general models to reduce effects such as
hallucinations~\cite{hu2023survey, pan2023unifying, wei2021knowledge,
yang2021survey, yang2023chatgpt}. Example schemes include Knowledge Graph
Prompting (KGP)~\cite{wang2023knowledge}, Graph Neural Prompting
(GNP)~\cite{tian2023graph}, Think-on-Graph (ToG)~\cite{sun2023think}, Knowledge
Solver (KSL)~\cite{feng2023knowledge}, KnowledGPT~\cite{wang2023knowledgpt}, and
others~\cite{brate2022improving, luo2023reasoning}.
Zhu et al.~\cite{zhu2023llms} discuss how LLMs can be used for enhancing KG
construction and tasks.
Wen et al.~\cite{wen2023mindmap} present MindMap, a framework to perform
reasoning on KG data. Pertinent triples from a KG are retrieved and the LLM is
prompted to answer a question based on these triples and show the reasoning
process by generating a ``mind map'' in the form of a textual reasoning tree.

Retrieval Augmented Generation (RAG) enhances the abilities of LLMs by enabling
the retrieval of documents into the LLM context to provide more accurate and
relevant responses. MRAG~\cite{besta2024multi} focuses on the multi-aspect
problems where as structure-enhanced RAG schemes employ different strategies for
structuring text to improve retrieval quality. A common idea is to construct a
knowledge graph from text, which enables retrieval amongst entities and
relationships~\cite{jiang2024hykge,delile2024graph,hussien2024ragbased,bui2024crossdata,xu2024retrieval}.
RAPTOR~\cite{sarthi2024raptor} generates multi-level summaries for clusters of
related documents, building a tree of summaries with increasing levels of
abstraction to better capture the meaning of the text.
Graph RAG~\cite{edge2024from} creates a knowledge graph, and summarizes
communities in the graph, which provide data at the different levels of
abstraction.
%

\subsection{Challenges and Future Directions}

Despite their numerous benefits, knowledge graphs also pose several challenges.
Their heterogeneity, as mentioned earlier, is one such challenge. It requires
the development of sophisticated techniques for KG embedding that can
effectively capture and preserve the diverse structures and semantics inherent
in the knowledge graphs~\cite{oag_heterogeneous_kgs}.

Another challenge lies in the dynamic nature of knowledge. As new information becomes available, KGs need to be updated to reflect this new knowledge. This requires efficient methods for knowledge graph updating and evolution~\cite{dynamic_kgs_embedding}.

Furthermore, the quality of the knowledge graph is heavily dependent on the quality of the input data. Hence, ensuring the accuracy and reliability of the data used to construct the KG is another significant challenge~\cite{kg_2021_data, kg_data_quality_eval}.

The existing methods for generating knowledge graph embeddings still suffer several severe limitations. Many established methods only consider surface facts (triplets) of knowledge graphs. However, additional information, such as entity types and relation paths, are ignored, which can further improve the embedding accuracy. The performance of most traditional methods that do not consider the additional information is unsatisfactory. Recently, some researchers have started to combine additional information with a knowledge graph to improve the efficiency of embedding models~\cite{Artif-Intell-Rev}.

Finally, more efficient processing of KGs is of great relevance, in the face of the ongoing growth of the dataset sizes. One strategy for achieving more performance is incorporating hardware acceleration techniques.


\section{Overview of Hardware Acceleration}

Hardware acceleration involves offloading specific computational tasks from the CPU to specialized hardware components within a system, leading to more efficient task processing.
There are several kinds of hardware acceleration, including GPU (Graphics
Processing Unit) for graphics and parallel tasks, DSP (Digital Signal Processor)
for handling signals like audio, FPGA (Field-Programmable Gate Array) which can
be customized for different uses after its production, ASIC (Application-Specific Integrated Circuit) designed for specific tasks, and NPU or AI Accelerators aimed at speeding up machine learning tasks. Figure~\ref{fig:overview} presents an overview of the covered hardware acceleration areas for knowledge graph processing.

To facilitate hardware acceleration, various technologies like CUDA by NVIDIA for general-purpose computing on GPUs, OpenCL for programming diverse systems, DXVA for hardware-accelerated video decoding, and WebGL for web-based graphics are used. These tools enable improved performance and energy efficiency, freeing up CPU resources for other tasks and enhancing the user experience. However, challenges such as compatibility issues, increased software complexity, higher initial costs, and the risk of hardware failure also arise. Despite these challenges, the benefits of hardware acceleration, including better performance and efficiency, make it a key element in advancing technology, especially as we move towards more specialized processing tasks.

\section{GPUs \& Knowledge Graphs}

Graphics Processing Units (GPUs) have emerged as powerful units for a multitude of computationally intensive tasks. Initially made for gaming graphics, GPUs are now crucial in many devices including smartphones, computers, and gaming consoles. Moreover, GPUs support massive parallelism, making them ideal for tasks that require heavy computation, such as machine learning, scientific computing, and cryptocurrency mining. They are now a key component in supercomputers and data centers. Nine out of the ten top supercomputers in the TOP500 list are powered by GPUs~\cite{top500}.

\subsection{Fundamental GPU Concepts}

GPUs have several some key benefits that make them suitable for accelerating
knowledge graph applications. GPUs support \textbf{massive parallelism} and
excel at performing many operations simultaneously with thousands of cores,
making them ideal for processing large-scale knowledge graphs, executing graph
algorithms, and training machine learning models on knowledge graphs. Efficient
\textbf{memory management} of GPUs can greatly speed up the processing of
knowledge graphs.
GPUs can use \textbf{thread-level parallelism} and employ \textbf{warp
scheduling}. Specifically, GPU threads are grouped into warps, which are scheduled for
execution together. By carefully organizing threads that access adjacent graph
nodes or edges into the same warp (a technique known as ``coalescing''), one can
maximize memory access efficiency and minimize warp divergence, leading to
significant performance improvements. GPU frameworks often allow the
\textbf{asynchronous execution} of different operations, enabling the overlap of
computation and memory transfer to hide latencies and improve throughput.
Additionally, \textbf{stream prioritization} of operations can ensure responsive
interactive querying in knowledge graph applications.
\textbf{Hardware-accelerated libraries} such as cuGraph~\cite{cugraph} (from
NVIDIA's RAPIDS suite) provide GPU-accelerated graph analytics algorithms, which
can be used in knowledge graph applications. For deep learning on knowledge
graphs, libraries such as PyTorch Geometric offer GPU-accelerated graph neural
network layers.

\subsection{Knowledge Graphs with GPUs}

GPUs have been increasingly leveraged in knowledge graph applications due to
their capability for parallel processing in two main areas. First, knowledge
graphs are typically stored and indexed using high-performance graph database
engines. Data analytics and machine learning techniques are applied to data
stored in graph databases by extracting relevant data from the knowledge graphs
using query languages (SPARQL for RDF graphs and Cypher, Gremlin as well as
others for property graphs). A large body of work~\cite{gpu-database-survey,
graph-gpu-survey} has focused on accelerating
graph databases and individual knowledge graph queries using GPUs.
%
%
Second, knowledge graphs are employed to learn inductive information
using either supervised or unsupervised machine learning approaches. Researchers
have focused on the use of GPUs for accelerating this learning process which we
outline below.

\subsubsection{Knowledge Graph Embeddings}




Embedding matrices are large and they typically do not fit into the limited GPU
memory. A common approach to address this challenge is to keep the embeddings in
the main memory and transfer them to the GPU memory as needed. However, this
results in severe latency penalties if GPUs exchange data with the main memory
frequently.

In a parallel setting where multiple workers together train a model, the graph and the embeddings need to be partitioned across workers.
Depending on the partitioning strategy, workers might need access to
embeddings of entities and relations that are not local to them. Workers also need to synchronize their updates to the embedding matrices.
Depending on the sampling strategy employed, workers might require further remote access for entities of negative triples.
This results in a high degree of communication between workers, which can be a bottleneck in the training process.

Several works have explored parallel training of knowledge graph embedding models on GPUs, tackling the above challenges.
%
\textbf{DGL-KE}~\cite{dgl-ke} is a distributed training framework for KGE models that uses a hybrid CPU-GPU system. It employs a distributed key-value store for both the knowledge graph structure and the embeddings, using shared CPU memory.
A GPU worker unit retrieves embeddings from CPU memory, updates them, and then
writes the embeddings back to the CPU memory.
In DGL-KE, the knowledge graph is partitioned via METIS~\cite{metis} such that
most of the entity and relation embeddings are local, in order to minimize communication between the compute units.
%
GPUs are not efficient in handling random memory access, and hence DGL-KE samples the negative triples in the CPU, and then transfers them to the GPU for training. 
This sampling is done from the local METIS partition to ensure that there is no increase in remote accesses for negative samples.
Other optimizations in DGL-KE include relation partitioning and overlap of gradient updates for relation as well as entity embeddings.

\textbf{GraphVite}~\cite{graphvite} focuses on multi-GPU training of KGE models.
In line with past works, it tackles the challenges of limited GPU memory and bus
bandwidth as well as synchronization overhead. GraphVite partitions the
embeddings in a way that avoids CPU-GPU or inter-GPU communication during
training. Positive triples are partitioned so, that they access pairwise
disjoint embeddings, and negative triples are sampled from the local partition.
Workers perform mini-batch updates on the local embeddings and synchronize only at the end of an epoch.

\textbf{Marius}~\cite{marius} is a framework designed for efficient computation of graph embeddings on a single machine by leveraging partition caching and buffer-aware data orderings to minimize disk access and interleave data movement with computation. The pipeline updates node embedding parameters in CPU memory asychronously, allowing for staleness, while the relation embeddings are updated in GPU memory synchronously. This design choice is based on the observation that updates to node embedding vectors are sparse, whereas updates to relation embedding parameters are dense due to the smaller number of edge-types in real-world graphs.

Another way to use GPUs for KGEs is to transform the knowledge graph completion problem into a similarity join problem, which can be efficiently processed by GPUs. This method can leverage the metric properties of some KGE models, such as TransE~\cite{bordes2013translating} and RotatE~\cite{sun2019rotate}, to reduce the number of vector operations and filter out irrelevant candidates. By using GPUs, this method can achieve fast and accurate knowledge graph completion on large-scale datasets~\cite{lee2023fast}.

\subsubsection{Graph Neural Networks}


A graph neural network (GNN)~\cite{besta2024demystifying} is a neural network in which input samples are graphs. GNNs, as opposed to embeddings, support end-to-end learning. Thus, GNNs can be used to
solve various KG-related tasks like link prediction, knowledge graph alignment,
and node classification~\cite{kg-gnn-survey}. GPUs
can accelerate the computation of GNNs by exploiting the parallelism and
locality of graph operations and utilizing the specialized hardware features of
GPUs. For example, \textbf{TC-GNN}~\cite{tc-gnn} proposes the use of tensor cores to accelerate sparse
matrix multiplication in GNNs by transforming the sparse graph data into dense
tensors. \textbf{PiPAD}~\cite{pipad} proposes to use pipelining and parallelism techniques to improve the efficiency and scalability of dynamic GNN training on GPUs.

Traditional KGE methods mainly focus on predicting the legitimacy between two entities and a particular relation type. GNNs have been shown to be effective in capturing the topological features of entities such as shapes of neighborhood sub-graphs which are overlooked by the traditional KGE methods~\cite{rgcn}.
However, their model complexity is higher in terms of the number of trainable parameters.
%

\textbf{Sheikh et al.}~\cite{gnn-kge} propose three key strategies to scale GNNs to
large knowledge graphs. Their system leverages vertex-cut partitioning to create self-sufficient graph sections and employs local negative sampling within partitions, significantly reducing communication overhead. It also utilizes edge mini-batch training, allowing efficient handling of large graph sections on GPUs.

\textbf{TinyKG}~\cite{tinykg} is a  a memory-efficient framework for training
Knowledge Graph Neural Networks. Traditional training of these networks is memory-intensive due to the need to store all intermediate activation maps for gradient computation, making deployment challenging in memory-constrained environments. TinyKG addresses this by using exact activations during the forward pass and storing a quantized version in the GPU buffers. During the backward pass, these quantized activations are dequantized for gradient computation. TinyKG employs a simple quantization algorithm to compress activations, reducing the training memory footprint with minimal accuracy loss.

\subsubsection{Symbolic Learning and Rule Mining}

Machine learning techniques that learn numerical models are hard to interpret
and quantify. Instead, symbolic learning can be used to learn hypotheses in a
logical (symbolic) language that ``explains'' sets of positive and negative edges.
Such hypotheses are interpretable and quantifiable (e.g., ``all airports are
domestic or international''), partially addressing the out-of-vocabulary issue.
Symbolic learning techniques such as rule and pattern mining are used to discover
interesting patterns over knowledge graph data. GPUs can exploit features of rule mining algorithms for more performance, for example frequent itemset generation, candidate pruning, support counting, and confidence evaluation. There exists a large body of work that focuses on accelerating rule mining specifically, and frequent itemset mining more generally, using GPUs. Most of them focus on accelerating the underlying algorithm like Apriori~\cite{apriori} that is widely used for rule mining. We refer the reader to a recent survey~\cite{fim-survey} of several GPU accelerated frequent itemset mining solutions proposed in literature for further reference.

\subsubsection{Graph Analytics and Mining}


The application of analytical methods to large-scale graphs is known as graph analytics. Such algorithms frequently examine the graph topology, or how nodes and groups of nodes are related. GPUs can accelerate a wide range of graph algorithms such as breadth-first search (BFS), single-source shortest path (SSSP), and community detection~\cite{graph-gpu-survey}.
%

\textbf{WikiSearch}~\cite{kgkeywordsearch} is an efficient parallel keyword
search engine designed for large-scale knowledge graphs, with a focus on the
Wikidata Knowledge Base, though it is applicable to other knowledge graphs as
well. To exploit parallelism, a novel approach for keyword search based on the
central graph method is proposed. Unlike traditional methods that approximate
the group Steiner tree problem, this approach can naturally operate in parallel
and returns compact, information-rich answer graphs. It is optimized for both multi-core CPU and GPU architectures. WikiSearch also introduces a novel pruning strategy based on keyword co-occurrence to refine search results further.

\ifconf
\else
\textbf{Scalable Multi-hOp REasoning (SMORE)}~\cite{smore} is a framework for
both single-hop knowledge graph completion and multi-hop reasoning on large
knowledge graphs, that involves predicting answers
\fi
to queries that span multiple relations or hops in the graph, which requires capturing complex dependencies and performing logical operations over entities and relations. The computational complexity for such tasks increases significantly with the number of hops, leading to higher memory requirements and processing times.
For a massive knowledge graph containing hundreds of millions of entities, it is not feasible to materialize training instances, and training data needs to be efficiently sampled on the fly with high throughput to ensure efficient utilization of computation resources. 
SMORE addresses this with a novel bidirectional rejection sampling approach for efficient online training data generation and an asynchronous system design that overlaps data sampling, embedding computation, and CPU-GPU communication. 
Also, graph partitioning is not feasible for multi-hop reasoning, as it requires traversing multiple relations in the graph, which will often span across multiple partitions.
SMORE is designed to operate in a shared memory environment, bypassing the limitations of graph partitioning in multi-hop reasoning, and demonstrates near-linear speed-up with the number of GPUs used for training.

In the field of biomedicine, the connections among various biomedical entities,
including drugs, diseases, symptoms, proteins, and genes play a crucial role in
understanding the underlying mechanisms of diseases and drugs. Biomedical
knowledge graphs play an important role in representing these connections and
are used in various applications such as drug discovery and repurposing.
\textbf{Distributed Accelerated Semiring All-Pairs Shortest Path
(DSNAPSHOT)}~\cite{dsnapshot} is a scalable knowledge graph analytics system
that can perform all-pairs shortest path (APSP) computation on large biomedical
knowledge graphs. It exploits the relation between the semiring
GEMM~\cite{semiring-gemm} and the APSP computation, and implements a
GPU-optimized distributed semiring GEMM kernel, the key operation in the Floyd-Warshall algorithm for APSP computation. Further, DSNAPSHOT proposes optimizations for both inter-node and intra-node communication, and achieves 90\% parallel efficiency on the Summit supercomputer.

\subsubsection{Graph Visualization}

Due to high dimensionality, heterogeneity, and sparsity of data, displaying
knowledge graphs might be difficult. By offering parallel computing capability,
high memory bandwidth, and specialized hardware characteristics for graphics
tasks, GPUs can make it possible for large-scale graph data to be rendered and
processed more quickly, thereby speeding up knowledge graph visualization. One
such example is \textbf{KG4Vis}~\cite{kg4vis}, a knowledge graph-based approach for visualization recommendation. It uses a TransE-based embedding technique to learn the embeddings of both entities and relations of the knowledge graph from existing dataset-visualization pairs. Such embeddings intrinsically model the desirable visualization rules and can be accelerated by GPUs.

\subsection{Challenges \& Limitations}

As knowledge graphs continue to grow in size and complexity, GPUs will likely play an increasingly important role in managing and extracting value from these datasets. However, GPUs also have several disadvantages that might make them unsuitable for some knowledge graph applications.

The \textbf{cost} of high-performance GPUs can be steep, which can be a barrier
to their use, especially for small organizations or individual developers. GPUs
have their own onboard memory, which is typically much less than the main memory
available to a CPU. While this memory is typically faster, the \textbf{limited
memory capacity} can be a challenge when working with large datasets that do not
fit into the GPU's memory. GPUs are more \textbf{complex to program} than CPUs.
Writing code that effectively leverages the parallel processing capabilities of
a GPU can require a different approach than what many developers are accustomed
to~\cite{bulavintsev2021adaptation}. GPUs often use more power and generate more
heat than CPUs, which can lead to additional \textbf{hardware requirements}
regarding power supply and cooling in a computer system. Not all tasks can be
effectively parallelized and see benefits from a
GPU~\cite{bulavintsev2019flattening}. Tasks with heavy data dependencies or
those that are inherently sequential may not see a performance improvement on a
GPU, and might even be slower than on a CPU. Such \textbf{limited tasks} may
benefit from applying a hybrid (CPU+GPU) processing
strategy~\cite{bulavintsev2017gpu}.


Hence, it is worth noting that not all knowledge graph tasks can benefit from GPU
acceleration. Certain operations, such as graph updates or graph schema
modifications, may not be well-suited for GPU parallelism. The effectiveness of
GPU acceleration will ultimately depend on specific graph algorithms, data
sizes, and hardware configurations. Thus, concrete benchmarking and
microarchitectural analysis of various knowledge graph-related tasks is required to understand the degree to which each task can benefit from GPU acceleration.

\section{FPGAs \& Knowledge Graphs}

Field-Programmable Gate Arrays (FPGAs) have emerged as integral components in contemporary digital electronics, facilitating the development of custom digital circuits with a degree of versatility unmatched by other devices. Characterized by arrays of programmable logic blocks and configurable interconnects, FPGAs offer a distinctive combination of adaptability and performance.

\subsection{Fundamental FPGA Concepts}

The cornerstone of any FPGA, \textbf{logic elements} and \textbf{logic blocks}
comprise arrays of both combinational and sequential circuit elements.
Programmable in nature, they can be tailored to execute a myriad of logical
functions, laying the groundwork for the vast functionalities FPGAs are known
for. The pathways of \textbf{configurable interconnects} serve a pivotal role in
an FPGA's architecture, facilitating signal routing across the device. Their
adaptability ensures seamless communication between discrete segments of a given
design, optimizing the device's functionality. A quintessential aspect of an
FPGA's reprogrammability is its \textbf{configuration memory}. This component
retains the user-defined design logic, effectively determining the FPGA's
operational behavior. Acting as the interface between the FPGA and its external
environment, \textbf{I/O blocks} are instrumental in the device's ability to
both send and receive signals, thereby ensuring effective communication with
other devices or components. In light of the stringent timing constraints often
associated with FPGA applications, effective clock distribution and management
are paramount. Mastery over \textbf{clock management} is crucial for the
successful deployment of FPGA-based designs. Analogous to programming languages
in the realm of software development, \textbf{Hardware Description Language
(HDL)} such as VHDL and Verilog as well as, more recently, other
abstractions such as \textbf{High-Level Synthesis (HLS)}~\cite{definelicht2018designing} are employed to define and describe
circuit behavior within an FPGA.

\subsection{Knowledge Graphs with FPGAs}

FPGAs have been increasingly utilized for knowledge graph processing due to
their high parallelism and configurability, which can significantly enhance the
efficiency of graph computations. FPGAs can be used to accelerate various graph
processing tasks, including defect detection in software
code~\cite{yan2021knowledge}, high-throughput updates on dynamic
graphs~\cite{wang2021grasu}, efficient traversal of edge-labeled directed
graphs~\cite{10020406} and automated composition and execution of Semantic Web
queries~\cite{werner2015automated}. FPGAs have also been used for
implementing various graph algorithms, such as BFS or
PageRank~\cite{heidari2018scalable, gui2019survey, besta2019graph}.


Different techniques have been employed to optimize the use of FPGAs for
knowledge graph processing. For instance, \textbf{GraSU}~\cite{wang2021grasu},
an FPGA library designed for the Xilinx Alveo™ U250 accelerator card, exploits
the spatial similarity of graph updates to improve overall efficiency. GraSU
outperformed two state-of-the-art CPU-based dynamic graph systems significantly
in terms of update throughput. Another technique involves the use of a pipeline
approach that combines parallel BFS and nondeterministic finite automaton for
efficient graph traversal~\cite{10020406}. Additionally, the use of partial
runtime-reconfiguration enables transparent query evaluation on an
FPGA~\cite{werner2015automated}.


Several specific schemes and mechanisms have been developed to optimize the use
of FPGAs for knowledge graph processing. For instance, a work-stealing-based
scheduler, \textbf{HWS}~\cite{agostini2020balancing}, has been designed to
optimize workload balance on heterogeneous CPU-FPGA systems. Another example is
the implementation of a stochastic matrix function estimator on FPGAs to boost
the performance and energy efficiency of subgraph centrality
computations~\cite{giefers2016energy, giefers2016analyzing}. Furthermore, an accelerator for quantized
Graph Convolutional Networks (GCNs) with edge-level parallelism has been
developed, using low-precision integer arithmetic during
inference~\cite{yuan2022qegcn}, which demonstrated significant speedups and
energy savings compared to other models.

\subsection{Advantages \& Disadvantages}

FPGAs offer several advantages for semantic knowledge graph processing. They
provide high parallelism and configurability, which can significantly enhance
the efficiency of graph computations~\cite{yan2021knowledge}. FPGAs can also
provide significant speedups and energy savings compared to other
models~\cite{yuan2022qegcn}. Despite the advantages offered by FPGAs and their
rapid growth, the use of FPGA technology is restricted to a narrow segment of
hardware programmers due to their code written differently using a hardware
description language to design the FPGA configuration. The challenge with the
HDL approach is that configuring an FPGA requires both coding skills and a
detailed knowledge of the underlying hardware, and the required expertise is not
widely available. More recent abstractions such as HLS~\cite{definelicht2018designing, definelicht2019hlslib, definelicht2020flexible, matteis2020fblas, johnson2022temporal, definelicht2022fast}
attempt to alleviate these issues.
Additionally, while FPGAs can provide significant performance
improvements for certain tasks, they may not benefit all
queries~\cite{werner2015automated}.

\section{Processing-In-Memory \& Knowledge Graphs}



Processing-In-Memory (PIM) is a promising way to alleviate the data movement bottleneck~\cite{mutlu2022modern, ghoseibm2019}, i.e., the waste of execution cycles and energy due to moving data between memory/storage and compute units, in current processor-centric computing systems (e.g., CPU, GPU).

\subsection{Fundamental PIM Concepts}

There are two main PIM trends. 
The first one is called \textbf{Processing-Near-Memory (PNM)} and consists of placing compute logic near the memory arrays (e.g., DRAM subarrays, banks, ranks)~\cite{boroumand2021google, besta2021sisa, nair2015active, cho2020mcdram, denzler2023casper, deoliveira2021IEEE, syncron}.
\textbf{Processing-Using-Memory (PUM)} is the other one, which leverages the analog operational properties of memory components (e.g., cells, sense amplifiers) to perform computation~\cite{orosa2021codic, xi2020memory, girard2020survey, seshadri.micro17, hajinazarsimdram}.
PIM represents a successful research trend in recent years, and several commercial PIM systems and prototypes~\cite{gomez2023evaluating, item2023transpimlibispass, diab2023aim, giannoula2022sigmetrics, gomezluna2022ieeeaccess, niu2022isscc, lee2022isscc, ke2021near, lee2022improving, lee2021hardware, kwon202125} have been presented. 


\subsection{Knowledge Graphs with PIM}

Graph Neural Networks use deep learning to process graph data, including
knowledge graphs~\cite{besta2024parallel}. GNNs can solve different knowledge
graph tasks, such as link prediction, knowledge graph alignment and reasoning, and node classification~\cite{kg-gnn-survey}.
%
There are several classes of GNNs: GCNs~\cite{kipf2016semi}, attentional GNNs~\cite{velivckovic2017graph} and
message-passing GNNs~\cite{bresson2017residual}.
\textbf{GCNs} are composed of several GCN layers, each computing two steps: the
aggregation of vertex features (a \textit{reduce} operation), and a combination
of features (an \textit{update} operation with typically fully-connected
layers). After the loss computation, the backward pass is composed of
feature/weight gradients computation (update), and feature gradients aggregation
(reduce).
While update operations (e.g., matrix multiplication) are compute-bound, reduce operations (e.g, gather-reduce-scatter) are very memory-bound. As such, reduce operations are good candidates for PIM-based acceleration. 
%
Several recent works~\cite{huang2022practical, mandal2022coin, yoo2023sgcn, huang2022accelerating, zhou2021gcnear} propose PIM acceleration for GCNs. 
Some of these works deploy PUM techniques such as ReRAM-based crossbars~\cite{mandal2022coin, huang2022accelerating}. 
Other works use PNM techniques with processing units in DDR DIMMs~\cite{zhou2021gcnear, huang2022practical} and in HBM stacks~\cite{yoo2023sgcn}.

PUM approaches rely on crossbar arrays, which help minimizing data movement in
reduce operations, and computing matrix multiplication efficiently.
\textbf{ReFlip}~\cite{huang2022accelerating} proposes a unified crossbar-based
PUM architecture that supports both compute-bound and memory-bound kernels. With
software/hardware co-optimizations, ReFlip maps both types of kernels efficiently onto the massive parallelism of its ReRAM-based crossbar arrays.
\textbf{COIN}~\cite{mandal2022coin} targets the huge communication overheads of
GCNs. For example, processing the Nell~\cite{carlson2010toward} knowledge graph causes 2.7TB of data moving between nodes of a baseline crossbar-based PUM architecture. COIN proposes an optimized on-chip interconnection network for efficient communication between compute elements and between the crossbars inside each compute element. The network design is applicable to different crossbars such as ReRAM- and SRAM-based, but COIN prefers ReRAM, which is significantly more energy efficient.

PNM approaches combine heterogeneous computing units that are specialized for different steps.
\textbf{GNNear}~\cite{zhou2021gcnear} integrates an ASIC with matrix multiply and vector processing units and PNM-enabled DIMMs. Update operations are computed on the ASIC, while execution units in the buffer chip of the DIMMs compute reduce operations.
Huang et al.~\cite{huang2022practical} tackle the large memory footprint and
data movement needs of GNNs with memory pooling. The authors propose a
customized memory fabric interface for low-latency and high-throughput
communication across PNM units in memory extension cards. The PNM units contain
a RISC-V core, a matrix multiply unit, and a vector processing unit.
\textbf{SGCN}~\cite{yoo2023sgcn} exploits the sparse nature of intermediate GCN
features to reduce the memory footprint (via compression) and optimize communication. SGCN places aggregation units (with SIMD MACs) and combination units (with a systolic array for matrix multiplication) near HBM memory.

\subsection{Advantages and Disadvantages}
PUM approaches for GNNs offer the advantage that the same crossbar-based PUM unit can accelerate both memory-bound and compute-bound kernels. Their main disadvantage is that they are based on memory technologies that are not yet mature (e.g., limited endurance and high area of ADCs in ReRAM and other non-volatile memories). 

PNM approaches tailor their execution units to the specific needs of each step,
which represents an advantage of their approach. However, they have yet to show how their performance would scale for knowledge graphs exceeding their memory capacity.


While the aforementioned works show great promise for GCN acceleration, their evaluations are all based on simulation. 
We hope to see soon efficient implementations of GNNs on existing real-world PIM architectures~\cite{gomez2023evaluating, item2023transpimlibispass, diab2023aim, giannoula2022sigmetrics, gomezluna2022ieeeaccess, niu2022isscc, lee2022isscc, ke2021near, lee2022improving, lee2021hardware, kwon202125} and future ones.

\section{Cluster-level RDMA \& Knowledge Graphs}

Remote Direct Memory Access (RDMA) is a mechanism for achieving high performance and scalability in both the supercomputing as well as the cloud data center landscapes~\cite{recio2007remote, gerstenberger2013enabling, mitchell2013using, wang2015hydradb, kalia2014using, simpson2020securing, huang2012high, islam2012high, lu2013high, woodall2006high, poke2015dare, liu2004high, kalia2016design, besta2015active, besta2014fault, schmid2016high}.
RDMA has grown popular as RDMA-enabled network interface cards have become widely used, and is commonly supported in modern  interconnects~\cite{besta2020high, besta2020fatpaths}. 
Overall, RDMA has many use-cases, particularly in distributed environment. Examples include speeding up data replication~\cite{burke2021prism, jha2019derecho, kim2018hyperloop, taleb2018tailwind, zamanian2019availability}, transactions~\cite{Wei_RDMA_HTM_transaction_paper, besta2023demystifying, Dragojevic_farm_paper, wei2018drtmh, Zamanian_rdma_transaction_paper}, index queries~\cite{ziegler2019rdmaindex}, file systems~\cite{di2022building}, general queries~\cite{Binnig_reldb_paper, Rodiger_reldb_1_paper, di2019network}, or analytical workloads~\cite{barthels2015rdmajoin, rdma_reldb_paper, besta2024parallel, besta2023demystifying, besta2023gdi, besta2023graph, strausz2022asynchronous, besta2020communication, solomonik2017scaling, besta2017push, besta2015accelerating}.

\subsection{Fundamental RDMA Concepts}

In general, the advantages of RDMA stem from the fact that communication bypasses the OS and the CPU, reducing or eliminating overheads such as interrupts.
While one can harness RDMA in different ways, highest performance is usually achieved with \emph{fully-offloaded one-sided communication}. In this approach, processes communicate by directly accessing dedicated portions of other processes' memory. In the established one-sided communication specification included in the Message-Passing Interface~\cite{mpi3}, this portion is called a \emph{window}.

One-sided accesses are done with communication operations referred to as \emph{put}s and \emph{get}s. They -- respectively -- write to and read from windows, offering very low latencies and most often outperform other communication paradigms such as message passing~\cite{gerstenberger2013enabling}.
Other useful RDMA operations include remote \emph{atomics} such as Compare-and-Swap or Fetch-and-Add~\cite{besta2015accelerating, schweizer2015evaluating, mpi3, Herlihy:2008:AMP:1734069} that are often accelerated by the interconnect hardware. They enable very fast fine-grained synchronization. To enforce data consistency between windows, operations called \emph{flushes} are employed to explicitly synchronize memories.
The communication operations come in two variants, \emph{blocking} (operation execution blocks till completion) and \emph{non-blocking} (operation execution returns immediately upon initiating communication). The latter can additionally increase performance by overlapping communication and computation~\cite{gerstenberger2013enabling}, with the user taking responsibility to synchronize memories at some point after starting the call. All of these routines are supported by most RDMA architectures.





\subsection{Knowledge Graphs with RDMA}

\textbf{Cray Graph Engine (CGE)}~\cite{maschhoff2015porting, rickett2018loading} is a system developed by Cray to support executing very large-scale RDF triple stores on top of Cray high-performance computing systems.
CGE's design is based on the Partitioned Global Address Space (PGAS) abstraction, in which one creates a single logical memory pool encompassing all the physical distributed memories over all compute nodes. Thus, CGE effectively implements the Single Program Multiple Data (SPMD) model and uses a purely one-sided RMA programming model, where memory access is treated as effectively uniform (i.e., at the programming level, one does not distinguish between local or remote accesses). Hence, one does not need to consider problems such as efficient graph partitioning.
Simultaneously, to achieve high performance, CGE heavily relies on different hardware features offered by the targeted systems. Such features include latency hiding, passive parallelism, and high network throughput for small remote requests that occur commonly in the targeted graph workloads. For example, one of the architectures used in the evaluation, the XMT2 system, used the Threadstorm processors with 128 hardware streams per node. For the whole system of a typical size (64--512 nodes), this amounts to a total of thousands to tens of thousands of software threads that can be used in a given single graph query. Finally, the CGE implementation relies on a low-level high-performance networking communication library~\cite{ten2010dmapp} by Cray. 

Improvements for graph analytics queries~\cite{maschhoff2017quantifying} with CGE were made by the use of non-blocking communication for large exchanges and by processing intermediate solutions in stages to exploit locality.
Additional recent improvements~\cite{rickettoptimizing} make CGE more portable by replacing low-level DMAPP operations with one-sided MPI routines, as well as simplifying the software stack. Several optimizations are employed to enable container performance matching that of native execution. Message-aggregation within each compute node is employed for join/scan/merge operations to reduce the number of messages. That
number is further reduced by improving the communication patterns to enable storing of messages for the same compute node consecutively. Distinct flushes for put and get operations improve the overlap of computation and communication.

\textbf{A1}~\cite{buragohain2020a1} is a distributed in-memory graph database developed by Microsoft. It adds graph abstraction and query engine layers on top of an improved FaRM key-value store~\cite{Dragojevic_farm_paper}. FaRM already comes with transaction support and uses one-sided read and write operations as well as RPCs to implement its functionality. A1 further improves FaRMs transactions by employing an optimistic multi-version concurrency control scheme by introducting a global clock and timestamps. RDMA is implemented with the RoCEv2 protocol~\cite{infiniband2014rocev2} and DCQCN is used for congestion control. RDMA-based unreliable datagrams are used for clock synchronization and leases. A1 is latency-optimized by employing data structures, which reduce the number of read operations, co-locating data likely to be requested at the same time like nodes and their edges as well as RPC aggregation to reduce the number of messages. A1 uses a semi-structured data model based on Bond~\cite{microsoft2016bond} with strictly-typed edges and weakly-typed nodes.

\textbf{Wukong}~\cite{shi2016fast} is a research-oriented distributed in-memory RDF triple store. Its storage layer is implemented using a simplified version of a RDMA-friendly distributed hashtable based on DrTM-KV~\cite{Wei_RDMA_HTM_transaction_paper}. Wukong duplicates edges during graph partitioning to store self-contained subgraphs on each compute node to preserve locality. It supports indices based on type and predicate. These indices are treated as a special kind of nodes and are usually replicated. Strings are stored separately and mapped to unique IDs to reduce network bandwidth. Wukong supports concurrent query execution with full history pruning, data (in-place) and/or execution (fork/join) migration as well as task stealing for load balancing and to reduce query latency. Originally providing limited update support, Wukong spawned several improved implementations. Wukong+S~\cite{zhang2017sub} adds support for stream queries as well as incremental key-value updates. Wukong+G~\cite{wang2018fast} uses GPUs to further improve query throughput by using the GPU memory as cache as well as the massive compute power of GPUs for triple parsing. Adaptive query scheduling was further proposed~\cite{yao2021wukong+} for Wukong+G to combine the processing of multiple queries similar to the fusion of kernels.

\textbf{RDMA\_Mongo}~\cite{huang2019rdma}, a document-oriented NoSQL database, uses RDMA writes to replace part of its TCP/IP-communication layer. \textbf{Nessie}~\cite{cassell2017nessie}, a key-value store, uses cuckoo hashing with RDMA for its key-value operations. Nessie decouples index and data storage to improve locality. \textbf{HERD}~\cite{kalia2014using}, another key-value store, uses one-sided RDMA writes and two-sided RDMA send/receives to complete each of its operations with a single network round trip. The key-value store \textbf{HydraDB}~\cite{wang2015hydradb} uses RDMA to accelerate its read operations and key caches with timestamps to reduce network pressure for highly skewed workloads. Similarly, \textbf{InnerCache}~\cite{yang2016innercache} uses one-sided RDMA to accelerate reads from the key-value store acting as an application cache and two-sided semantics for writing data. RDMA-based Memcached has been used for the integration of the Hadoop storage layer (HDFS) with the underlying high performance parallel filesystem Lustre to improve the I/O performance of big data analytics~\cite{islam2015accelerating}. Additionally a non-blocking API extension for Memcached to improve communication and computation overlap as well as an enhanced runtime design for hybrid use with SSDs was proposed~\cite{shankar2016high}.

Finally, the \textbf{Graph Database Interface (GDI)}~\cite{besta2023gdi} has recently been proposed to deliver a toolbox for designing a scalable and high-performance data access and transaction layer for general graph databases that can also be used to maintain knowledge graphs. Its RDMA-based implementation, \textbf{GDI-RMA}~\cite{besta2023graph}, has been shown to scale to more than a hundred thousand compute cores and to label- and property-rich graphs with more than 500 billion edges. The key mechanisms used for high performance are one-sided non-blocking RDMA communication, hardware-accelerated network atomic operations, and collective communication, a form of group communication that has been tuned over decades by the MPI community~\cite{mpi3}.

\subsection{Advantages and Disadvantages}

RDMA usually enables significant performance advantages in terms of both latency and bandwidth. The former is enabled by eliminating expensive parts of the communication pipeline (such as interrupts) and by supporting features such as network-accelerated atomic operations. The latter is facilitated by features such as the overlap of computation and communication. On the other hand, RDMA is usually more complex to program and maintain. This, however, has been alleviated with efforts such as the one-sided communication within MPI~\cite{mpi3, gerstenberger2013enabling} or by GDI~\cite{besta2023gdi, besta2023graph}.



\section{Future Research Opportunities}


We identified four research gaps in current solutions, that provide
opportunities for future studies: scalability, energy efficiency, real-time
processing and the integration with other technologies.

While current hardware solutions have shown promise in handling large-scale
knowledge graphs, there is a pressing need to address the \textbf{scalability}
challenges posed by the ever-growing size and complexity of semantic data.
The energy consumption of hardware accelerators, especially when processing
extensive knowledge graphs, remains a concern. Research into more
\textbf{energy-efficient} hardware designs is crucial.
The ability to process and update knowledge graphs in \textbf{real-time},
especially in dynamic environments, is still an area with limited research.

The synergy between different forms of hardware acceleration and \textbf{other
technologies} is not fully explored. Existing works, such as
DaCe~\cite{ben2019stateful, ben2023bridging}, focus on effective and efficient
execution of different workloads on the underlying diverse hardware, targeting
-- among others -- machine learning~\cite{rausch2022data}, linear algebra
kernels~\cite{kwasniewski2021parallel}, and -- more recently --
GNNs~\cite{bazinska2023cached}. Extending this line of works towards knowledge
graphs specifically is an interesting research opportunity. Another, related,
opportunity is to combine this approach with other emerging technologies, such
as quantum computing or neuromorphic computing.

In conclusion, while significant strides have been made in the domain of
hardware-accelerated semantic knowledge graph processing, there remains a vast
landscape of uncharted territory. By addressing the identified research gaps and
capitalizing on the highlighted opportunities, the scientific community can pave
the way for more efficient, scalable, and innovative solutions in the future.

\subsection{Novel Hardware Acceleration Schemes}

Beyond the hardware acceleration techniques previously mentioned, there are several innovative hardware solutions that can enhance knowledge graph processing.

\textbf{SmartNICs} are advanced network interface cards (NICs) that allow the
processing of tasks on the NIC~\cite{hoefler2017spin} instead of the CPU. For load balancing in
distributed settings, SmartNICs can distribute incoming queries to different
servers to ensure efficient utilization of resources. They can also be used in
the context of data preprocessing, where SmartNICs preprocess and filter
irrelevant data, therefore speeding up the data ingestion process of the
knowledge graph.

Originally designed for machine learning, \textbf{Tensor Processing Units
(TPUs)} can also be harnessed for knowledge graph tasks that incorporate machine
learning. TPUs can accelerate the training and inference of GNNs to be used for
node and graph classification as well as link prediction in knowledge graphs.
TPUs can also compute efficiently node and edge embeddings for similarity
searches and clustering in knowledge graphs.

Additionally various \textbf{AI accelerators}, such as Google's Edge TPU and
Intel's Nervana Neural Network Processor, can process knowledge graphs in order
to detect anomalies or inconsistencies to guarantee data integrity. Google's
Edge TPU can be used on edge devices to locally update a knowledge graph as new
data streams in to ensure that graph processing remains current.

\textbf{Quantum Computing} is an emerging field with the potential to redefine
computing. For knowledge graph processing it holds the promise of more efficient
computation for various algorithms such as graph isomorphism, i.e. determining
if two graphs are structurally identical, or pathfinding (finding the shortest
path or optimal connections between nodes).

Inspired by the human brain, \textbf{Neuromorphic Computing} can be advantageous
for traditionally challenging tasks. Neuromorphic chips can help
to identify patterns or trends in knowledge graphs and can aide in tasks like
recommendation systems or predictive analytics. Similar as a human brain learns
from experience, neuromorphic computing can adaptively learn from data in the
knowledge graphs and refine queries and results over time.

Utilizing superconducting circuits that function at ultra-low temperatures,
\textbf{Cryogenic Computing}, though in its infancy, has potential for
large-scale knowledge graph processing tasks. At extremely low temperatures,
superconducting circuits can process vast amounts of data simultaneously, which
allows for massive parallel processing of large-scale knowledge graph analytics,
where multiple queries and computations are performed concurrently. Cryogenic
computing can also offer significant energy savings, making the processing more
sustainable and cost-effective.

\textbf{Chiplet} architectures~\cite{chiplets-status-challenges}, where multiple
silicon dies are integrated into a single package, have gained significant
traction in the chip design industry~\cite{chiplet-book}. Chiplets come with a
wide range of benefits, including modularity, reusability, flexibility,
specialization, cost-efficiency and reduced time-to-market. Even though we are
not aware of any chiplet-based accelerators specifically designed to process
knowledge graph, chiplets have played an important role in knowledge graph
processing. A team of researchers was elected as Gordon Bell
Prize~\cite{gordon-bell-prize} finalists for running their COAST
(communication-optimized all-pairs shortest path)~\cite{exascale-biomedical}
algorithm for knowledge graphs on Frontier, the world's first exascale
supercomputer, which uses chiplet-based AMD EPYC CPUs~\cite{amd-exascale,
amd-chiplets}. There is also a variety of propositions for chiplet-based
accelerators for general graph processing~\cite{tascade, waverscale-chiplets,
popstar, popstar-eval}, which hints for a large potential of leveraging their
modularity and cost-efficiency for hardware accelerators tailored to knowledge
graph processing.

As the number of compute cores in modern processors and accelerators is steadily
increasing, networks-on-chips (NoCs) have emerged as a scalable \textbf{On-Chip
Interconnect} solutions. In a NoC, data packets are sent through a series of
links and routers, akin to computer networks~\cite{packets-not-wires}. The de
facto standard topology for NoCs is a 2D mesh~\cite{survey-noc-proposals,
survey-routing-mesh}, however, more elaborate topologies such as
Slim NoC~\cite{slimnoc} or sparse Hamming graphs~\cite{iff2023sparse} have been
proposed. Most accelerators for graph processing rely on these mesh
topologies~\cite{scalagraph, gnn-accelerator}, while some hardware architects
argue that a mesh is not suitable for graph algorithms, as these algorithms
often cause data movement between physically distant cores. One approach to
tackle this challenge is the use of small-world
networks~\cite{small-world-networks} as NoC topologies for graph processing
accelerators~\cite{graph-accelerator-small-world-1,
graph-accelerator-small-world-2}. We believe that there is significant value in
a thorough investigation of traffic caused by knowledge graph processing and a
subsequent evaluation of NoC topologies for knowledge graph accelerators.

Application-Specific Integrated Circuits (\textbf{ASICs}) mark a transformative
shift in VLSI design, enabling systems to be embedded within single chips rather
than assembled from multiple components~\cite{7141}. This evolution, akin to the
earlier microprocessor revolution, not only reshapes the electronics industry's
design and manufacturing strategies but also interconnects designers, CAE tool
developers, and ASIC vendors in intricate ways. Broadly, ASIC covers a spectrum
from programmable logic devices (PLD) to gate arrays (GA), standard cells (SC),
and full custom (FC) designs, with GA and SC being the most commonly referenced.
We find it surprising that there are no works on designing ASICs for knowledge
graphs, beyond the PIM-related works; it constitites a promising direction of
future development.

\section{Conclusion}
In this paper, we explore hardware acceleration for knowledge graph applications. We review the existing literature, identify main designs and trends in that area, benefits and drawbacks of hardware acceleration, as well as the challenges and opportunities for future research and development.
We consider GPUs, FPGAs, Processing-In-Memory, RDMA, and other forms of acceleration.
Our work can help design more efficient KG processing schemes.

 

\iftr
\section*{Acknowledgments}

%
This project received funding from the European Research Council (Project PSAP, No.~101002047), and the European High-Performance Computing Joint Undertaking (JU) under grant agreement No.~955513 (MAELSTROM). This project was supported by the ETH Future Computing Laboratory (EFCL), financed by a donation from Huawei Technologies. This project received funding from the European Union’s HE research and innovation programme under the grant agreement No. 101070141 (Project GLACIATION).
\fi

\bibliographystyle{IEEEtran}
\bibliography{bibliography}

\EOD
\end{document}